\documentclass[letterpaper,twocolumn,prb,showpacs,floatfix,superscriptaddress]{revtex4}
\usepackage{graphicx}
\usepackage{amsfonts,color}
\usepackage{amssymb,amsmath}
\usepackage{bm}

\newcommand{\be}{\begin{equation}}
\newcommand{\ee}{\end{equation}}
\newcommand{\bea}{\begin{eqnarray}}
\newcommand{\eea}{\end{eqnarray}}

\begin{document}

\title{Effects of Strong Correlations and Disorder in $d$-Wave Superconductors}

\author{Marcos Rigol}
\affiliation{Department of Physics, University of California, Santa Cruz, California 95064, USA}
\affiliation{Department of Physics, Georgetown University, Washington, District of Columbia 20057, USA}
\author{B. Sriram Shastry}
\affiliation{Department of Physics, University of California, Santa Cruz, California 95064, USA}
\author{Stephan Haas}
\affiliation{Department of Physics and Astronomy, University of Southern California, Los Angeles, 
California 90089, USA}

\date{\today}


\begin{abstract}

We use exact diagonalization techniques to study the interplay between strong correlations, superconductivity, 
and disorder in a model system. We study an extension of the $t$-$J$ model by adding an infinite-range 
$d$-wave superconductivity inducing term and disorder. Our work shows that in the clean case the magnitude of 
the order parameter is surprisingly small for low-hole filling, thus implying that mean-field theories might 
be least accurate in that important regime. We demonstrate that substantial disorder is required to destroy 
a $d$-wave superconducting state for low-hole doping. We provide the first bias free numerical results for 
the local density of states of {a strongly correlated $d$-wave superconducting model}, 
relevant for STM measurements at various fillings and disorders.
\end{abstract}

\maketitle

The combination of strong correlations and reduced dimensionality makes the theoretical understanding 
of high-$T_c$ superconductivity very difficult and consensus on its origin has not been 
reached.\cite{norman03,anderson07,scalapino06} Very recently, experiments using local probes, such as scanning 
tunneling spectroscopy (STS) and scanning tunneling microscopy (STM), have shown that the doped cuprates
are highly inhomogeneous\cite{expSTSSTM} (for a recent review, see Ref.\ \onlinecite{fischer07}). 
Specific aspects of high-$T_c$ superconductivity, such as the robustness of the  tunneling spectrum 
with respect to disorder,\cite{randeria_disorder} emphasize its contrast with more conventional disorder 
sensitive\cite{bcs_2} BCS-type superconductivity. 

In this work, we probe the effects of strong correlations and disorder in $d$-wave superconductors 
derived from a Mott insulator. We introduce and study a generalized  model derived from the $t$-$J$ 
model\cite{lee06} in which a superconducting (SC) ground state is argued to be inevitable. 
We consider a Hamiltonian ${\cal H}={\cal H}_{tJ} + {\cal H}_{d} + {\cal H}_{random}$, with
\begin{equation}
{\cal H}_{tJ}=-t\sum_{\langle i,j\rangle\,\sigma} 
\left[\tilde{c}^\dagger_{i\sigma}\tilde{c}^{}_{j\sigma} + \mathrm{H.c.} \right]
+ J \sum_{\langle i,j\rangle} \left[{\bf S}_i\cdot{\bf S}_j-\frac{1}{4}n_i n_j \right],
\label{tJ}
\end{equation}
where the sum $\langle i,j\rangle$ runs over nearest-neighbor sites, 
$\sigma = (\uparrow, \downarrow)$, and standard definitions for the projected 
creation and annihilation operators are employed.\cite{lee06}

The $t$-$J$ model is microscopically justified from either the one-band Hubbard model by a 
large $U$ expansion or more generally from a reduction of the three-band copper oxide model 
to a single-band model,\cite{3bandto1band} with a greater freedom for the parameter ratio 
$J/|t|$. While a mean-field theory (MFT) for the doped $t$-$J$ model\cite{mft,kotliar} gives 
a $d$-wave SC ground state, it is an uncontrolled approximation. In the MFT, there are often 
states with other broken symmetries in the proximity of the superconductor that can be missed. 
Most importantly, in view of the very strong constraint of single occupancy in the model, we 
expect significant quantum fluctuations, and the MFT cannot handle these precisely. Therefore 
it is not clear that the $t$-$J$ model {\em does} have a SC ground state for the ranges of 
parameters studied. In order to precipitate  a SC starting state, we add an attractive term
\begin{equation}
{\cal H}_d   = - \frac{\lambda_d}{L} \sum_{i,j=1}^L D^\dagger_i \; D_j
\label{dwave}
\end{equation}
where $D_i=(\Delta_{i,i+\hat{\textrm{x}}}-\Delta_{i,i+\hat{\textrm{y}}})$, $\Delta_{ij}=\tilde{c}_{i\uparrow}\tilde{c}_{j\downarrow}+
\tilde{c}_{j\uparrow}\tilde{c}_{i\downarrow}$, and $L$ is the number of lattice sites. 
This is  an infinite-range term  of the  type that BCS considered in their reduced 
Hamiltonian,\cite{bcs} while  building  in the $d$-wave symmetry of SC order. We have also 
considered imposing an extended $s$-wave symmetry, where the results are qualitatively quite different
and will be reported elsewhere. Within MFT, this model leads to the same $d$-wave state as found 
from the $t$-$J$ model.\cite{mft,kotliar} Our model is presumably a superconductor {\em for any} 
$\lambda_d \sim O(1)$ {in the thermodynamic limit}, and {\em for sufficiently large} $\lambda_d$, 
for any reasonable finite cluster. Notice that we have sidestepped the issue of the ``mechanism'' 
of superconductivity, which cannot be settled with studies of the kind undertaken here and focus 
instead on the nature of the state so produced. We argue below that despite the infinite-ranged 
nature of $H_d$, strong correlations produce a non-mean-field-like state; this state has an 
unexpectedly small order parameter (OP). 

Finally, we consider a quenched random disorder term of the form
\begin{equation}
{\cal H}_{random}   =  \sum_{i} \varepsilon_i n_i \label{disorder}
\end{equation}
where the $\varepsilon_i$'s are taken randomly from a uniform distribution between [$-\Gamma,\Gamma$]. 
The full  Hamiltonian Eqs.\ (\ref{tJ})-(\ref{disorder}) thus describes an 
inhomogeneous strongly correlated superconductor. In our study, we use numerical 
diagonalization of clusters with 18 and 20 sites. The dimension of the largest 
Hilbert space diagonalized here is $\sim 10^8$. 

\begin{figure}[!tb]
\begin{center}
  \includegraphics[width=0.44\textwidth,angle=0]{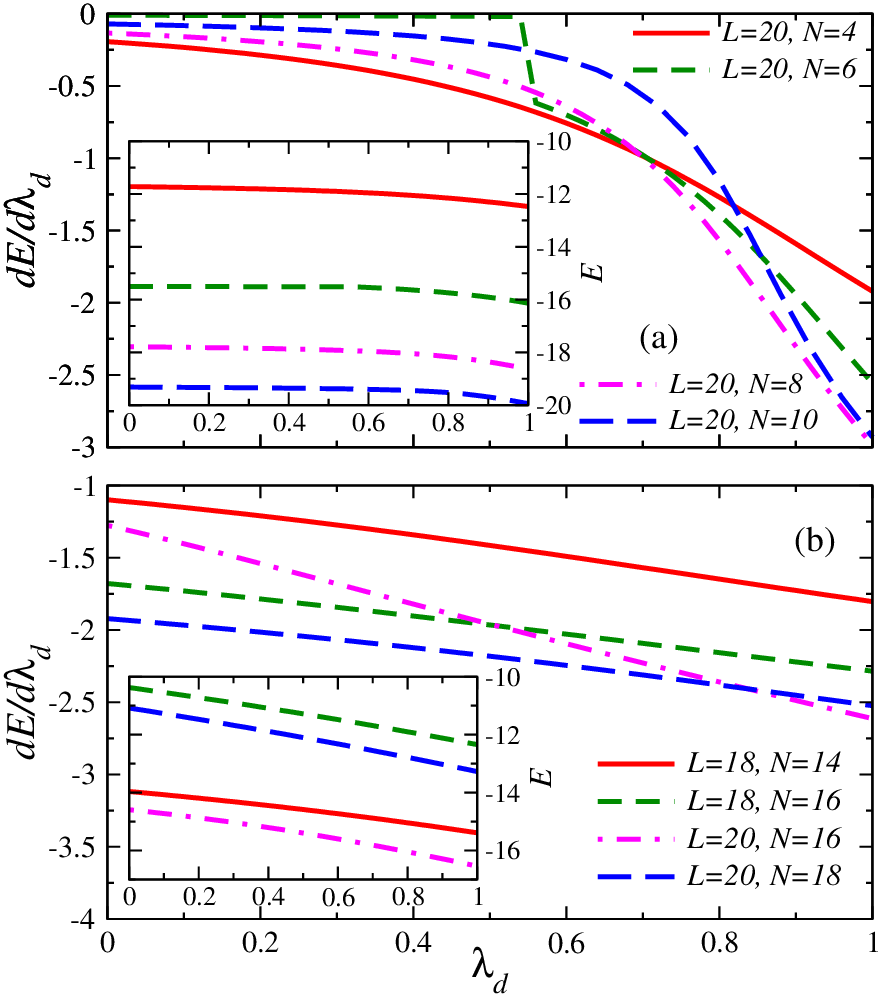}
\end{center}
\vspace{-0.6cm}
\caption{\label{EnergyvsLambda} (color online). Energy (insets) and its derivative 
(main panels) vs $\lambda_d$. (a) Low density of particles (less correlated) and (b) low 
density of holes (strongly correlated). $N$ is the number of electrons, and $J=0.3$. 
All energies are in units of $t$.}
\end{figure}

In our model, we are interested in  understanding how the evolution into the SC state occurs, 
as $\lambda_d$ is turned on. Towards this end, we show in Fig.\ \ref{EnergyvsLambda} the derivative 
of the energy (main panels) and the energy itself (insets) as $\lambda_d$ is increased, for different 
fillings of the 18 and 20 site clusters. For low fillings of electrons [Fig.\ \ref{EnergyvsLambda}(a)], 
we find that for certain cases  of the number of particles, level crossings  occur, as signaled by a 
jump in the energy derivative, indicating a change of the symmetry of the ground state.\cite{footnote_1} 
A similar jump is seen for {\it all fillings} between $N=4$ and $N=10$ in the 18-site cluster 
[not shown in Fig.\ \ref{EnergyvsLambda}(a)].  On the other hand, we find that for low-hole fillings 
in the 18 and 20 site clusters, the  energy derivative is continuous with $\lambda_d$, suggesting a 
particular compatibility between the $d$-wave order and the $t$-$J$ model. It is interesting that there 
is evidence of this compatibility from high-temperature expansions\cite{puttika} and exact diagonalization 
studies,\cite{dagotto} which are also unbiased such as the present one.

One of our diagnostic tools for studying the nature of the SC state is the $d$-wave pair density 
matrix $P_{ij}=\langle D^\dagger_i D_j\rangle$.\cite{odlro} Its largest ($\Lambda_1$) and next largest 
($\Lambda_2$) eigenvalues are computed and their ratio $R( \geq 1)$ is monitored. This ratio is an effective 
probe of the order, both for clean and disordered superconductors. Taking the ratio eliminates uninteresting 
normalization effects related to the change in the particle density, etc. This procedure, for example, eliminates 
the expected diminishing of all the eigenvalues of $P_{ij}$ as the hole doping decreases (due to Gutzwiller 
correlations). It is thus constructed as a pure number. For a SC ground state, it is expected to scale 
for large $L$ like $R\sim\Psi^2 L+\Phi$, where $\Psi$ is the dimensionless $O(1)$ OP (Ref.\ \onlinecite{odlro}) 
and $\Phi\sim O(1)$ represents the depletion (i.e., spillover) from the condensate. This depletion occurs 
due to  repulsive interactions, i.e., strong correlations. In the parallel case of a Bose system with 
$N_b$ bosons,\cite{leggett01} we expect $\Lambda_1 \sim O(N_b)$ while $\Lambda_2\sim O(1)$. 

\begin{figure}[!tb]
\begin{center}
  \includegraphics[width=0.44\textwidth,angle=0]{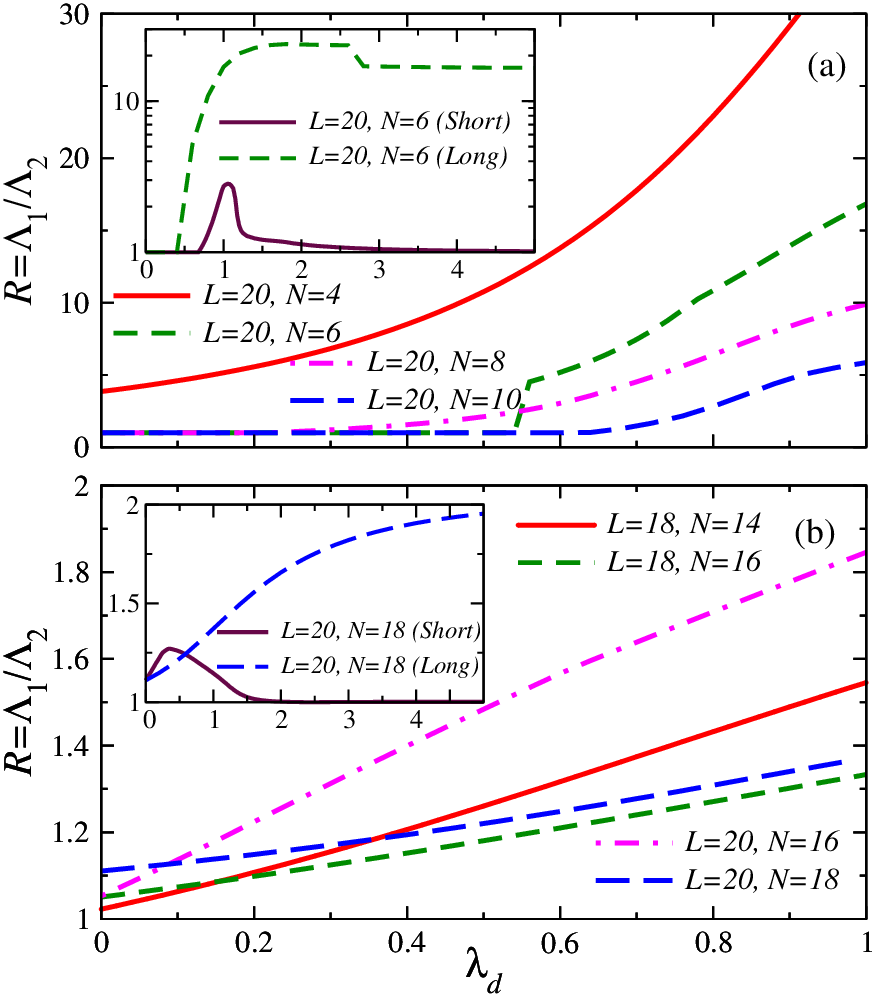}
\end{center}
\vspace{-0.6cm}
\caption{\label{OrderParametervsLambda} (color online). Ratio $R= \Lambda_1/\Lambda_2$  
with increasing $\lambda_d$.(a) Low density of particles (less correlated) and (b) 
low density of holes (strongly correlated). In the insets, we compare the infinite-range 
model Eq.\ (\ref{dwave})]  with a short-range version of  Eq.\ (\ref{dwave}) 
[taking $i=j$  and dropping  the  $L$ in the denominator] up to larger values of 
$\lambda_d$. $N$ is the number of electrons, and $J=0.3$.}
\end{figure}

In Fig.\ \ref{OrderParametervsLambda}, we show $R=\Lambda_1/\Lambda_2$ for different fillings of 
the 18 and 20 lattices as a function of $\lambda_d$. By comparing the low {\em electron}
filling [Fig.\ \ref{OrderParametervsLambda}(a)] to the low {\em hole} doping case 
[Fig.\ \ref{OrderParametervsLambda}(b)], one can gauge the effects of correlations for 
overdoped (a), and optimal or underdoped cuprates (b). For the lowest electron densities 
($N=4$), $R$ reaches very large values ($R>30$) and decreases as the density is increased, 
up to around $R\sim 6$ for ten electrons.  We notice that sometimes for low electron filling 
one needs $\lambda_d$ to exceed a critical value before $R$ starts increasing, 
{\it e.g.}, $N=6$ in Fig.\ \ref{OrderParametervsLambda}(a). This is a signature of a 
quantum phase transition into a SC state, and occurs in many but not all instances. 
The transition point coincides with the jump seen in the derivative of the energy in 
Fig.\ \ref{EnergyvsLambda}. Taken together, these confirm  that the new ground state has 
a different symmetry than the ground state of the plain vanilla $t$-$J$ model.

On the opposite end, for low-hole doping [two and four holes in 18 and 20 sites in 
Fig.\ \ref{OrderParametervsLambda}(b)], one can see that $R$ increases continuously with 
$\lambda_d$, i.e., no abrupt transition occurs. Figure \ref{OrderParametervsLambda}(b) 
also shows that in that regime $R$ increases very slowly with $\lambda_d$ and does 
not exceed $R=2$ for $\lambda_d\leq1$. We can interpret this as a small value of 
$\Psi$ and a large value of $\Phi$ as defined above, implying a large depletion 
of the condensate. This shows, that even in our infinite-range model, the SC OP at low doping 
is very strongly depleted. However by no means should we understand that {\em superconductivity} 
is weaker in that regime. From studies of purely bosonic systems, it is known that due to 
strong correlations,  the superfluid (SC) fraction (i.e., $\rho_s$) can be much larger than 
the condensate fraction (i.e., the OP $\Psi$) (see, {\it e.g.}, Ref.\ \onlinecite{bernardet02}). Correlations 
also tend to make superfluidity (superconductivity) more stable against perturbations. 

We now compare our results for the infinite-range model of Eq.\ (\ref{dwave}) to 
those produced by the more standard short-range case [Eq.\ (\ref{dwave}), 
for $i=j$ and no normalization by $L$ in the denominator] used in the literature 
dealing with the Hubbard model. The insets in Fig.\ \ref{OrderParametervsLambda} 
show that while in the infinite-range model $R$ saturates 
with increasing $\lambda_d$, in the short-range model $R$ attains a maximum value for 
$\lambda_d\sim1$ and then decreases towards unity. The latter occurs because for 
$\lambda_d\gg1$ one produces localized pairs, i.e., there is no long-range coherence.

\begin{figure}[!tb]
\begin{center}
  \includegraphics[width=0.44\textwidth,angle=0]{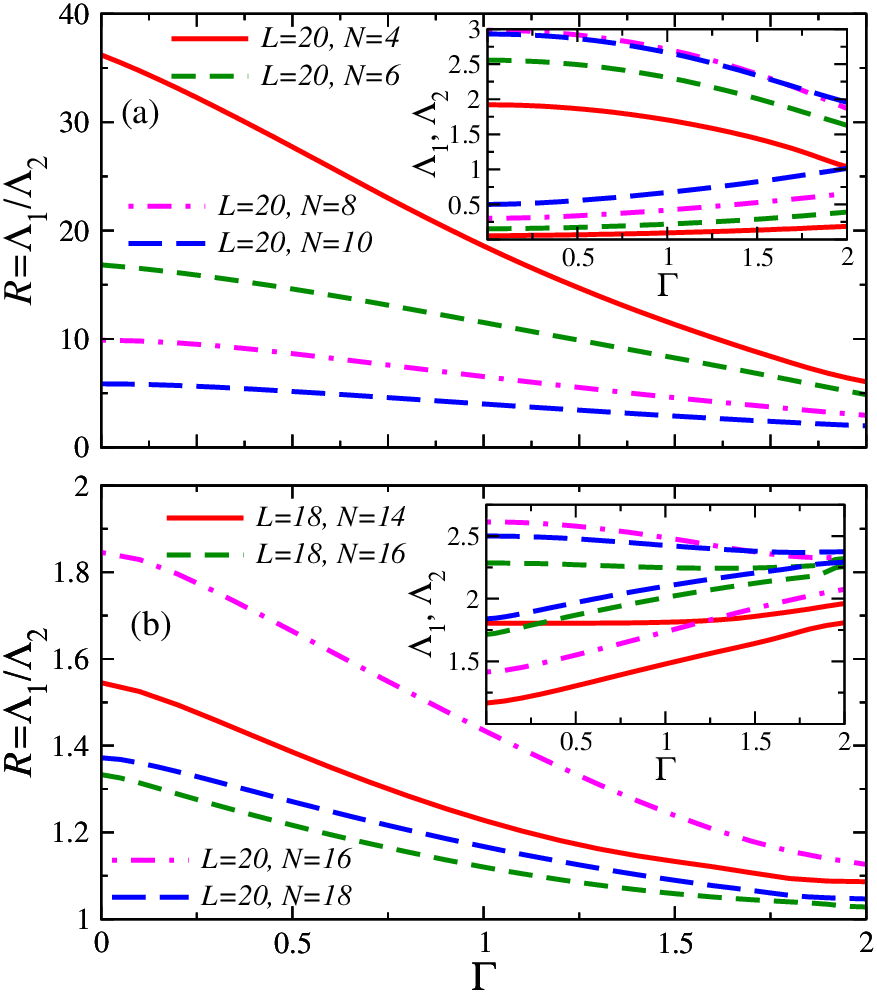}
\end{center}
\vspace{-0.6cm}
\caption{\label{OrderParametervsDisorder} (color online). Ratio $R= \Lambda_1/\Lambda_2$  
with increasing disorder ($\Gamma$). (a) Low density of particles (less correlated) and (b) 
low density of holes (strongly correlated). In the insets, we show the separate values 
of $\Lambda_1$ (upper plots) and $\Lambda_2$ (lower plots), corresponding to the results
shown in the main panels. Here $J=0.3$ and $\lambda_d=1$. These results were obtained 
averaging over ten different disorder realizations.}
\end{figure}

Turning to disorder, in the main panel in Fig.\ \ref{OrderParametervsDisorder}, 
we show how $R$ evolves with increasing disorder for $\lambda_d=1$. (Those results were obtained
averaging over ten different disorder realizations.)  Here one can see that disorder produces a 
very large reduction of $\Lambda_1/\Lambda_2$ for low electron  filling, i.e., disorder has 
a very large impact on the SC OP. For the case of low-hole 
density, the relative  reduction of $\Lambda_1/\Lambda_2$ is much smaller, i.e., as 
$\lambda_d$ had a small effect in increasing $R$, so is $\Gamma$ having a smaller 
effect in reducing it.

From the main panels in Fig.\ \ref{OrderParametervsDisorder}  we see that the effect of 
disorder in the SC state is always to make $R$ decrease. It is of considerable interests 
to understand what happens to $\Lambda_1$ and $\Lambda_2$ separately as $\Gamma$ is 
increased. Results for these quantities are presented in the insets in 
Fig.\ \ref{OrderParametervsDisorder}. There one can see that $\Lambda_1$ behaves 
qualitatively very differently between low-electron fillings and low-hole fillings. 
In the first case $\Lambda_1$ exhibits a very large reduction, which points 
towards the destruction of superconductivity. On the other hand, for low-hole doping,
$\Lambda_1$ is almost unaffected by the increase of disorder and can even be 
enhanced, as shown for 14 particles in 18 sites. Unexpectedly, the reduction of $R$ 
in this case is related to an increase of $\Lambda_2$.  This increases points towards a 
slower decay of $P_{ij}$ when disorder is increased. This suggests the possibility of an 
emerging algebraic long-range order, producing a different signature in the density matrix 
than the case of standard LRO. For example, in the 2D $XY$ model below $T_c$, or in the 
1D Heisenberg antiferromagnetic ground state, there is no true LRO, but several of the 
largest density matrix eigenvalues scale as $L^{\eta}$, with $\eta<1$. The system sizes we 
treat here are too small to make definitive statements.  However, it is interesting to note 
that for low-hole doping the behavior is qualitatively different from the low electron filling, 
in which the largest eigenvalue exhibits a large decrease with increasing disorder. 
An analysis of the data for the $s$-wave superconductor studied in Ref.\ \onlinecite{huscroft98} 
exhibits exactly the latter behavior, in contrast to the one we see for the 
SC $t$-$J$ model in the low-hole doping regime. Our results therefore suggest unusual 
power-law type superconductivity in the presence of disorder close to half filling. 

\begin{figure}[!tb]
\begin{center}
  \includegraphics[width=0.485\textwidth,angle=0]{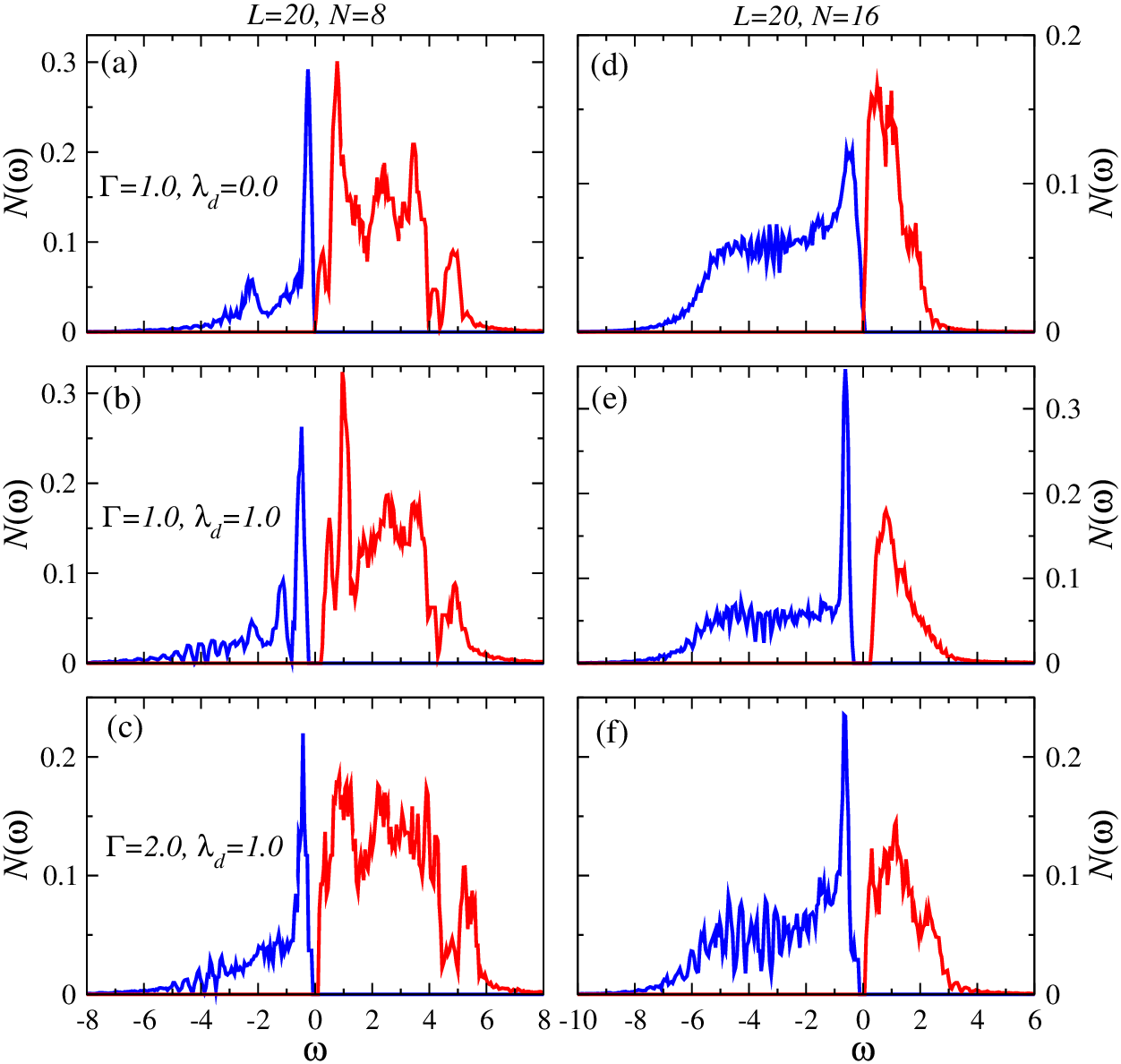}
\end{center}
\vspace{-0.6cm}
\caption{\label{STM} (color online). Averaged density of states 
[$N(\omega)$] for two different fillings ($N=8$ and $N=16$) of the 20 site 
cluster. We show results for: 
$\Gamma=1$, $\lambda_d=0$ (a) and (d); 
$\Gamma=1$, $\lambda_d=1$ (b) and (e); and 
$\Gamma=2$, $\lambda_d=1$ (c) and (f). $N(\omega)$ was computed as the average 
over different lattice sites and over two different disorder realizations. 
}
\end{figure}

In order to make connection with experimentally measurable STM curves, we show in Fig.\ \ref{STM} 
the local density of states of the $L=20$ site cluster for two different fillings in the presence 
of disorder and $\lambda_d$. Figures \ref{STM}(a) and \ref{STM}(d) correspond to fillings where the ground 
state of the plain $t$-$J$ model ($\lambda_d=0$) is adiabatically connected to the SC ground state 
at finite $\lambda_d$. In the presence of disorder, the density of states is  similar to the one 
reported previously for the translational invariant $t$-$J$ model with $L=16$ sites.\cite{dagotto94} 
{These curves display a striking asymmetry between adding a particle and 
taking out a particle, and the evolution of this asymmetry with doping is similar to that of the clean 
$t$-$J$ model.}

Adding the SC term ($\lambda_d>0$) to the disordered system opens a gap. This can be clearly seen 
in Figs.\ \ref{STM}(b) and \ref{STM}(e). Our system sizes are too small to see the V shape expected for a 
$d$-wave superconductor, i.e., we see a real gap. As disorder is increased, Figs.\ \ref{STM}(c) and \ref{STM}(f) 
show the reduction of the gap. From the results shown in Figs.\ \ref{STM}, we see that the SC gap closes 
only for a substantial disorder ($\Gamma \gtrsim 2 \lambda_d$). Our calculations therefore also sheds 
light on this aspect of the STM spectra, namely, the robustness against disorder.

In conclusion, we have presented and studied a variant of the $t$-$J$ model, with an infinite-range 
$d$-wave superconducting term. We have shown how the energy, its derivative, and the $d$-wave superconducting 
order parameter evolve with increasing the strength of the superconducting term. In addition to 
discontinuities in all the above quantities for low electron densities, we find a severe reduction of 
the magnitude of the order parameter at low-hole filling. This is a signature of strong 
quantum fluctuations near the Mott insulator. In relation to current STM experiments, we find that 
superconductivity survives considerable disorder close to half filling. The local density-of-states curves 
yield bias free (i.e., non variational) results for a strongly correlated $d$-wave superconductor 
in the presence of disorder and provide a picture of the large energy scale structure of this important object.

\begin{acknowledgments}

We acknowledge support from NSF under Contract NO. DMR-0706128 and DOE-BES under Contract NO. DE-FG02-06ER46319. 
We thank M. A. P. Fisher, G. H. Gweon, A. Pasupathy, M. Randeria, J. A. Riera, and R. T. Scalettar for helpful 
discussions. Computational facilities were provided by HPCC-USC center.

\end{acknowledgments}

\end{document}